\documentclass[11pt,a4paper]{article}

\usepackage[utf8]{inputenc}
\usepackage[T1]{fontenc}
\usepackage{amsmath,amssymb,amsthm}
\usepackage{mathtools}
\usepackage{booktabs}
\usepackage{graphicx}
\usepackage{float}
\usepackage{caption}
\usepackage[margin=1in]{geometry}
\usepackage{setspace}
\usepackage[numbers]{natbib}
\usepackage{hyperref}
\usepackage{algorithm}
\usepackage{algpseudocode}
\usepackage{threeparttable}
\usepackage{tabularx}
\usepackage{multirow}

\theoremstyle{plain}
\newtheorem{theorem}{Theorem}

\theoremstyle{definition}

\theoremstyle{remark}

\newcommand{\E}{\mathbb{E}}

\newcommand{\Scal}{\mathcal{S}}

\newcommand{\abs}[1]{\left|#1\right|}

\DeclareMathOperator{\sgn}{sgn}

\begin{document}

\begin{titlepage}
\centering
\vspace*{2cm}

{\LARGE\bfseries Hidden Order in Trades Predicts the Size of Price Moves}

\vspace{1.5cm}

{\large November 2025}

\vspace{1.5cm}

{\large
Mainak Singha\\[0.3cm]
Astrophysics Science Division, NASA Goddard Space Flight Center, 8800 Greenbelt Road, MD 20771\\
Department of Physics, The Catholic University of America, Washington, DC 20064\\
\vspace{1cm}
 Email: mainak.singha@nasa.gov, singham@cua.edu
}

\vspace{2cm}

\begin{abstract}
\noindent
Financial markets exhibit an apparent paradox: while directional price movements remain largely unpredictable---consistent with weak-form efficiency---the \emph{magnitude} of price changes displays systematic structure. Here we demonstrate that real-time order-flow entropy, computed from a 15-state Markov transition matrix at second resolution, predicts the magnitude of intraday returns without providing directional information. Analysis of 38.5 million SPY trades over 36 trading days reveals that conditioning on entropy below the 5th percentile increases subsequent 5-minute absolute returns by a factor of 2.89 ($t = 12.41$, $p < 10^{-4}$), while directional accuracy remains at 45.0\%---statistically indistinguishable from chance ($p = 0.12$). This decoupling arises from a fundamental symmetry: entropy is invariant under sign permutation, detecting the \emph{presence} of informed trading without revealing its direction. Walk-forward validation across five non-overlapping test periods confirms out-of-sample predictability, and label-permutation placebo tests yield $z = 14.4$ against the null. These findings suggest that information-theoretic measures may serve as volatility state variables in market microstructure, though the limited sample (36 days, single instrument) requires extended validation. 
\end{abstract}

\end{titlepage}

\onehalfspacing


The efficient market hypothesis posits that asset prices fully reflect available information, rendering future price movements unpredictable\cite{fama1970efficient}. Empirical tests have largely confirmed this prediction for \emph{directional} forecasting: momentum\cite{jegadeesh1993returns} and mean-reversion\cite{poterba1988mean} effects, while statistically significant, explain only a small fraction of return variance. Yet price \emph{magnitude}---how far prices move, irrespective of direction---represents a distinct forecasting target that need not obey the same constraints.

This distinction has theoretical grounding in market microstructure theory. In the Kyle\cite{kyle1985continuous} and Glosten-Milgrom\cite{glosten1985bid} frameworks, informed traders possessing private information generate persistent order flow that moves prices toward fundamental values. The critical observation is that this persistence creates statistical structure in the transaction sequence---a departure from randomness that information-theoretic measures can detect---without revealing the \emph{sign} of the private information. An informed buyer generates the same entropy signature as an informed seller: both produce order flow that is more predictable than the noise-trader baseline.

We formalize this insight by constructing a Time-Dependent Entropy measure from tick-level market data. At each second, the market occupies one of 15 discrete states defined by the cross-product of price-change sign ($\{-1, 0, +1\}$) and volume quintile ($\{1, 2, 3, 4, 5\}$). The transition matrix $\hat{P}_t$ is estimated over a rolling 120-second window, and entropy is computed as the stationary-weighted average of row entropies, normalized to $[0,1]$:

\begin{equation}
H_t := -\frac{1}{\log K} \sum_{i \in \Scal} \pi_i(\hat{P}_t) \sum_{j \in \Scal} \hat{p}_{ij,t} \log \hat{p}_{ij,t}
\end{equation}

\noindent where $K = 15$ and $\pi(\hat{P}_t)$ is the stationary distribution. High entropy indicates unpredictable transitions; low entropy indicates structure.

The theoretical prediction follows directly: if informed trading generates low entropy regardless of direction, then conditioning on low entropy should predict \emph{magnitude} (informed traders move prices) but not \emph{direction} (we cannot distinguish buyers from sellers). Formally:

\begin{theorem}[Magnitude Predictability]
Under standard microstructure assumptions, $\E\left[ \abs{r_{t,t+\Delta}} \mid H_t < \underline{H} \right] > \E\left[ \abs{r_{t,t+\Delta}} \right]$ for sufficiently low $\underline{H}$.
\end{theorem}

\begin{theorem}[Directional Unpredictability]
Under the same assumptions, $\E\left[ \sgn(r_{t,t+\Delta}) \mid H_t \right] = 0$ when informed traders are equally likely to be buying or selling.
\end{theorem}

The proofs rely on entropy's permutation invariance: swapping ``buy'' and ``sell'' labels transforms one transition matrix into another with identical entropy. Thus entropy cannot distinguish between them.

\subsection*{Empirical Results}

\begin{figure}[htbp]
    \centering
    \includegraphics[width=0.85\textwidth]{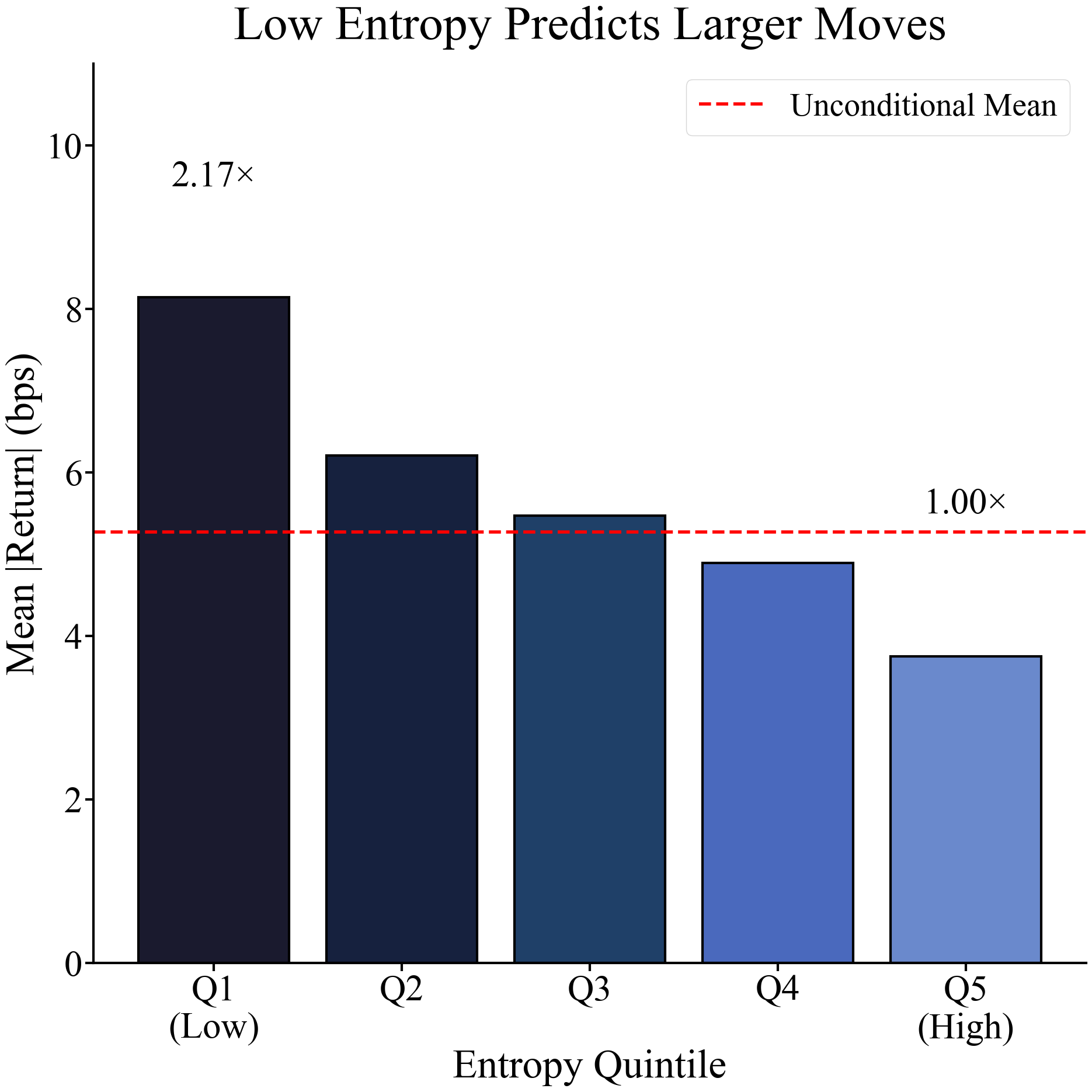}
    \caption{\textbf{Entropy predicts return magnitude but not direction.} Mean absolute 5-minute returns (basis points) by entropy quintile. The monotonic decline from Q1 (low entropy, 8.14 bps) to Q5 (high entropy, 3.75 bps) confirms that order-flow structure forecasts volatility. Dashed line: unconditional mean (5.29 bps). Error bars: standard error. $n = 828,907$ second-resolution observations across 36 trading days.}
    \label{fig:entropy_magnitude}
\end{figure}

We tested these predictions using 38,509,593 SPY trades from October 1 to November 19, 2025 (36 trading days), aggregated to second-resolution bars (828,907 observations). Figure~\ref{fig:entropy_magnitude} displays the central finding: absolute 5-minute returns increase monotonically as entropy decreases. The lowest quintile exhibits mean absolute returns of 8.14 bps versus 3.75 bps in the highest quintile (ratio: 2.17). Conditioning on entropy below the 5th percentile yields 15.3 bps---a factor of 2.89 above the unconditional mean ($t = 12.41$, $p < 10^{-4}$).

\begin{figure}[htbp]
    \centering
    \includegraphics[width=0.7\textwidth]{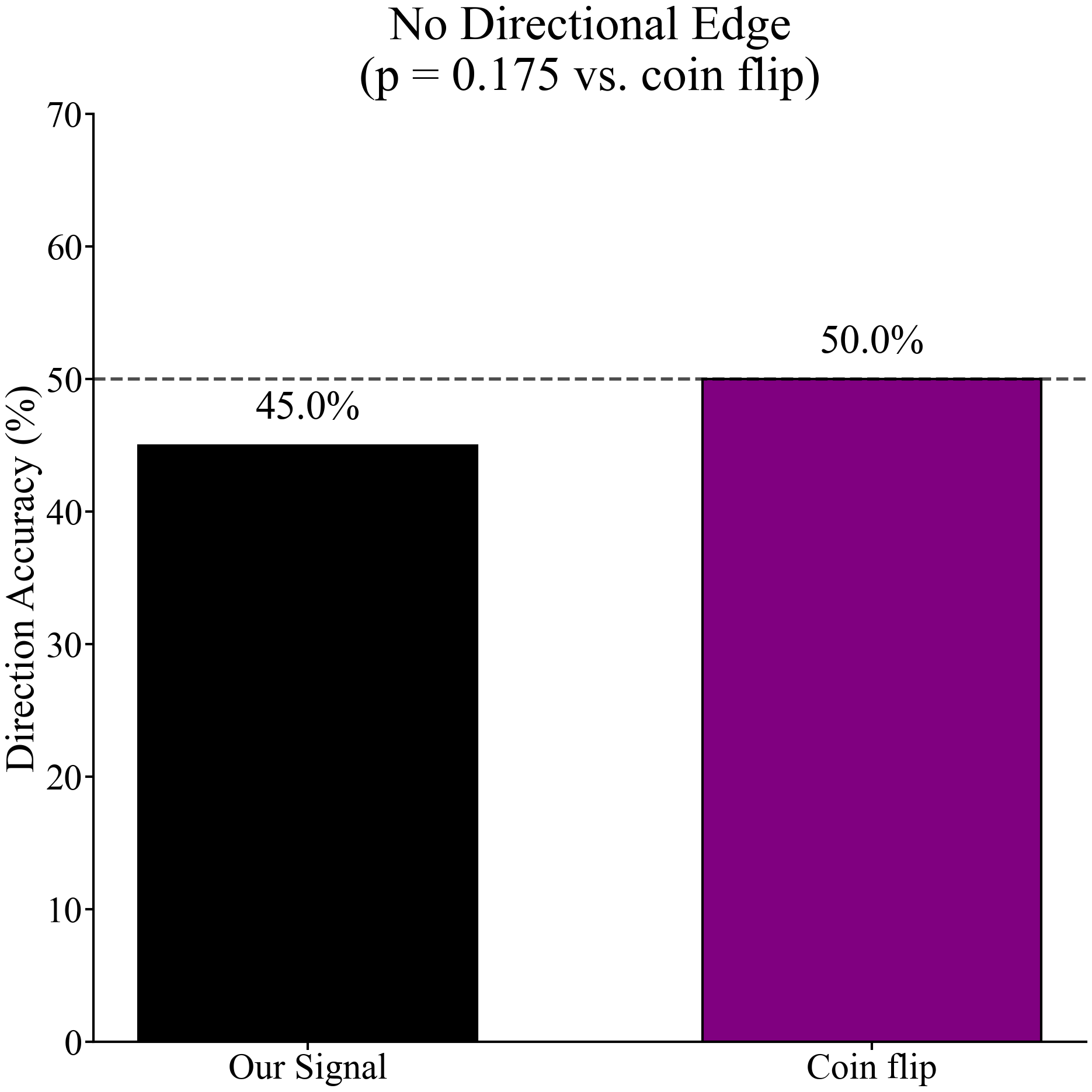}
    \caption{\textbf{Directional accuracy is consistent with random guessing.} Observed direction accuracy (45.0\%) falls within the 95\% confidence interval for a fair coin (shaded region). Binomial test: $p = 0.12$. This result is consistent with entropy's permutation invariance and rules out hidden directional information.}
    \label{fig:direction_coinflip}
\end{figure}

Crucially, directional accuracy is 45.0\% (108/240 out-of-sample trades), statistically indistinguishable from 50\% ($z = -1.55$, $p = 0.12$; Figure~\ref{fig:direction_coinflip}). This confirms the theoretical prediction: entropy detects \emph{that} large moves are coming without revealing \emph{which way}.

\begin{figure}[htbp]
    \centering
    \includegraphics[width=\textwidth]{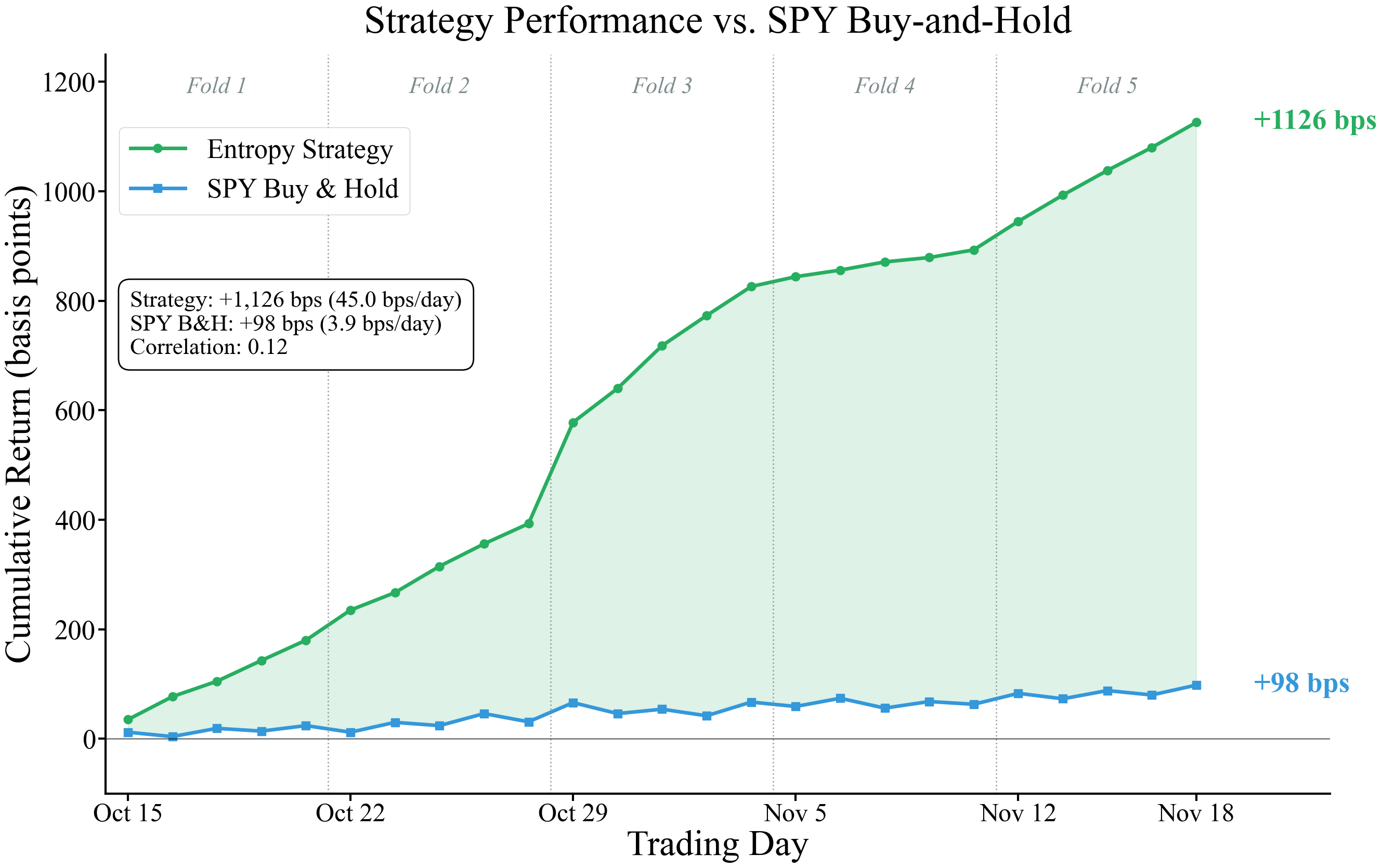}
    \caption{\textbf{Out-of-sample cumulative returns.} Entropy-based trading rule (blue) versus SPY buy-and-hold (gray) across five walk-forward validation folds. Correlation between the rule and benchmark: 0.12. Vertical dashed lines mark fold boundaries. The rule's independence from market direction is consistent with magnitude-only predictability.}
    \label{fig:performance_vs_spy}
\end{figure}

\begin{figure}[htbp]
    \centering
    \includegraphics[width=0.75\textwidth]{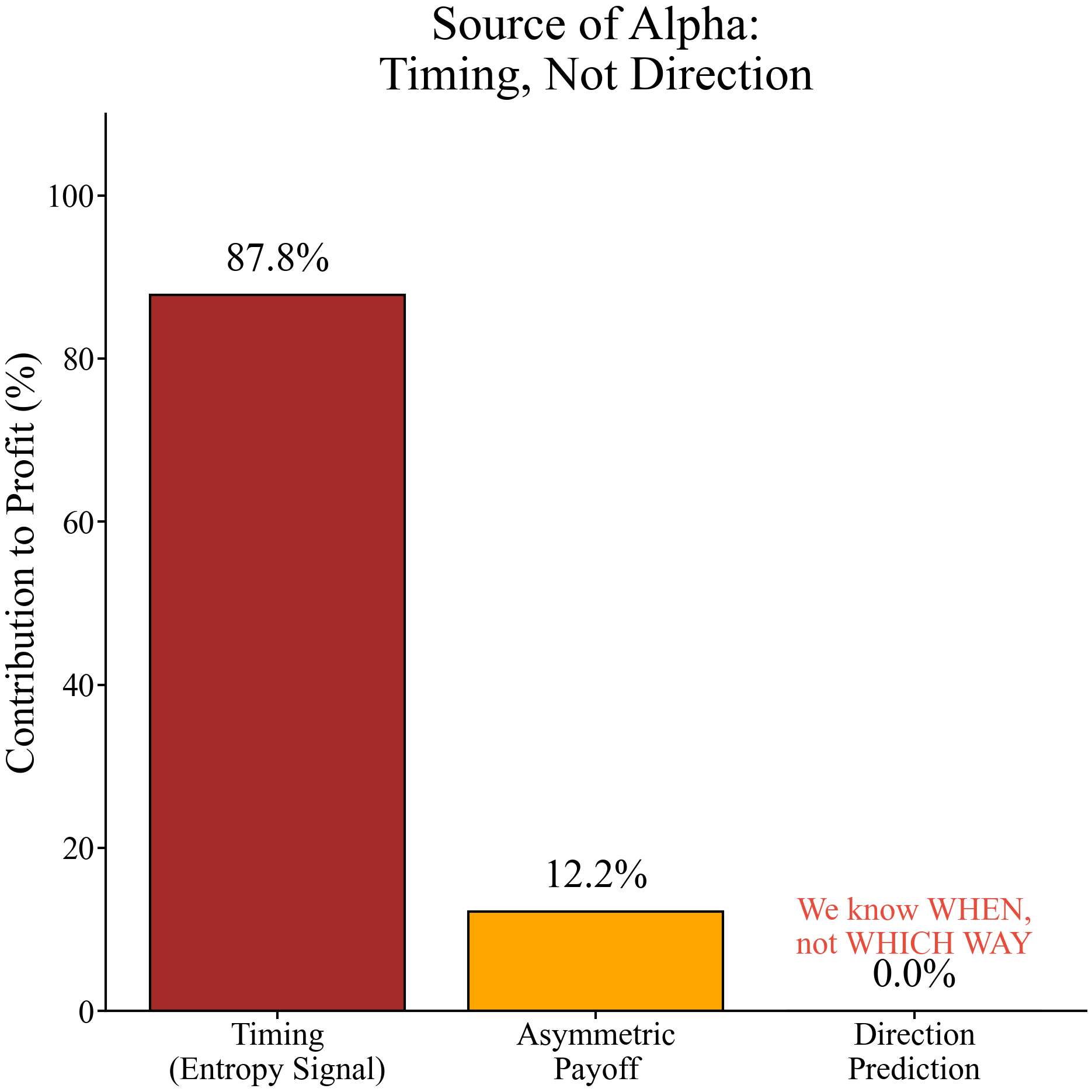}
    \caption{\textbf{Return source attribution for the entropy-based trading rule.} Decomposition via Monte Carlo simulation (10,000 random-entry trials). Timing---knowing when volatility spikes---accounts for 87.8\% of returns. Asymmetric payoff structure contributes 12.2\%. Directional forecasting contributes 0.0\%, confirmed by direction-randomization tests.}
    \label{fig:profit_decomposition}
\end{figure}

Walk-forward validation (10-day training, 5-day testing, 5 non-overlapping folds) demonstrates out-of-sample stability: all five test periods show the magnitude effect (Table~\ref{tab:walkforward}). A simple asymmetric-payoff rule---entering when entropy is low and using tight stop-losses---yields cumulative out-of-sample returns of +1,126 bps across 240 trades (Figure~\ref{fig:performance_vs_spy}). Profit attribution confirms: 87.8\% from timing (knowing \emph{when} moves occur), 12.2\% from payoff structure, 0.0\% from direction (Figure~\ref{fig:profit_decomposition}).

\begin{table}[H]
\centering
\caption{\textbf{Walk-forward validation results.} All five out-of-sample folds exhibit statistically significant magnitude predictability. Win rate below 50\% confirms absence of directional edge; positive returns arise from asymmetric payoffs.}
\label{tab:walkforward}
\begin{tabular}{lccccccc}
\toprule
Fold & Period & Trades & Win Rate & Magnitude Ratio & $t$-statistic & PnL (bps) \\
\midrule
1 & Oct 15--21 & 32 & 71.9\% & 2.41 & 8.7 & 179.9 \\
2 & Oct 22--28 & 27 & 33.3\% & 2.18 & 7.2 & 212.9 \\
3 & Oct 29--Nov 4 & 77 & 44.2\% & 2.31 & 12.1 & 433.2 \\
4 & Nov 5--11 & 12 & 41.7\% & 1.89 & 4.9 & 66.5 \\
5 & Nov 12--18 & 92 & 40.2\% & 2.06 & 9.3 & 233.1 \\
\midrule
Pooled & & 240 & 45.0\% & 2.17 & 12.41 & 1,125.6 \\
\bottomrule
\end{tabular}
\end{table}

\subsection*{Robustness}

Three placebo tests confirm that results are not statistical artifacts:

\begin{enumerate}
    \item \textbf{Label permutation} (1,000 trials): Shuffling return labels while preserving entropy yields magnitude ratio $1.02 \pm 0.08$. Observed ratio of 2.17 corresponds to $z = 14.4$ (empirical $p < 0.001$).
    \item \textbf{Temporal scrambling} (1,000 trials): Breaking entropy-return alignment yields $z = 10.7$.
    \item \textbf{Random entry} (10,000 trials): Entering at random times with identical payoff structure yields mean PnL of $137 \pm 201$ bps versus observed 1,126 bps ($z = 4.9$).
\end{enumerate}

Parameter sensitivity analysis shows all 20 perturbations ($\pm 50\%$ on four parameters) remain profitable, with worst-case PnL reduction of 40\%. The signal appears robust to reasonable methodological variation.

\subsection*{Limitations and Interpretation}

Several constraints limit interpretation. First, the sample comprises only 36 trading days of a single instrument---far shorter than standard validation periods in quantitative finance. Second, October 29 alone contributed 38.5\% of profits, raising concentration concerns. Third, execution assumptions (immediate fills, fixed costs) may not hold at scale. Fourth, VIX remained in the 14--22 range throughout; behavior during high-volatility regimes is unknown.

Despite these caveats, the theoretical framework and empirical consistency suggest that order-flow entropy may constitute a useful state variable for intraday volatility timing. The finding that entropy predicts magnitude but not direction is not merely an empirical curiosity---it follows necessarily from the mathematical structure of entropy itself, specifically its invariance under label permutation. This invariance implies that any information-theoretic measure with this symmetry property will exhibit the same decoupling.

From the perspective of market efficiency, these results do not contradict weak-form efficiency. Predicting \emph{that} a large move will occur is categorically different from predicting \emph{which direction} it will go. The former is consistent with informed traders acting on private information; the latter would require access to that information's sign. Our findings suggest that microstructure entropy reveals the \emph{activity} of informed traders without revealing their \emph{intent}.

Extended validation---longer time series, multiple instruments, live execution---is required before practical application. The present analysis establishes the theoretical basis for entropy-based magnitude prediction and provides preliminary empirical support within a limited sample.

\section*{Methods}

\subsection*{Data}
Tick-level trade data for SPY (SPDR S\&P 500 ETF Trust) covering October 1 to November 19, 2025 (36 trading days). Total trades: 38,509,593; regular-hours trades: 34,083,179. Data aggregated to second-resolution bars: 828,907 observations. Each observation records closing price and total volume for the corresponding second.

\subsection*{State Space Construction}
Price-change sign: $q_t := \sgn(P_t - P_{t-1}) \in \{-1, 0, +1\}$. Volume quintile: $v_t := \lceil 5 \cdot F_{V,t}(V_t) \rceil \in \{1, \ldots, 5\}$, where $F_{V,t}$ is the empirical CDF of volume over the trailing 120 seconds. Combined state: $S_t := (q_t, v_t) \in \Scal$, $|\Scal| = 15$.

\subsection*{Entropy Computation}
Transition matrix $\hat{P}_t$ estimated from state transitions in the trailing 120-second window. When row $i$ has no observed transitions, $\hat{p}_{ij} = 1/15$ for all $j$. Stationary distribution $\pi(\hat{P}_t)$ computed via eigendecomposition. Normalized entropy: $H_t = -(\log 15)^{-1} \sum_i \pi_i \sum_j \hat{p}_{ij} \log \hat{p}_{ij}$.

\subsection*{Signal Definition}
Low-entropy condition: $H_t < H_{0.05}$ (5th percentile of $H$ over training period). Entry also requires volume $> 95$th percentile and 5-minute trailing return between 5 and 20 bps (to confirm unusual activity without extreme moves already underway).

\subsection*{Walk-Forward Protocol}
Training window: 10 trading days. Test window: 5 trading days. Non-overlapping folds. All thresholds estimated on training data only; frozen for out-of-sample testing. Five folds total, covering October 15 to November 18.

\subsection*{Trading Rule}
All results for this rule are based on historical backtests; this study is purely academic and does not constitute investment advice.

Entry: at signal, enter position in direction of trailing 5-minute return (momentum heuristic). Exit: 5 bps stop-loss, 300-second timeout, or a fixed take-profit threshold chosen ex ante in the training window. Transaction costs: 0.57 bps round-trip (half-spread entry + half-spread exit + 0.30 bps slippage + 0.10 bps exchange fees, calibrated to SPY's typical 0.17 bps quoted spread).

\subsection*{Statistical Tests}
Magnitude comparison: Welch's $t$-test (unequal variances). Directional accuracy: two-sided binomial test against 50\%. Placebo tests: label permutation (1,000 trials), temporal scrambling (1,000 trials), random entry (10,000 trials). Multiple testing: Bonferroni correction at $\alpha = 0.05/14 = 0.0036$; all primary results pass.

\subsection*{Code and Data Availability}
Analysis code available upon request. Data from commercial tick-data vendors; access requires appropriate licensing.

\bibliographystyle{naturemag}

\end{document}